\documentstyle[12pt,amsfonts,amsbsy]{article}
\def\fmref#1{(\ref{#1})}

\def\vec#1{{\boldsymbol #1}}
\def\diGamma{\gamma}
\begin{document}
\title{High temperature QCD and QED with unstable excitations
\thanks{Work supported by GSI. GSI-Preprint 95-4 (1995), subm. to
  Nucl.Phys. {\bf B}}}
\author{P.A.Henning and R.Sollacher
\thanks{Electronic mail: P.Henning@gsi.de, sollach@tpri6c.gsi.de}\\
Theoretical Physics,
        Gesellschaft f\"ur Schwerionenforschung GSI\\
        P.O.Box 110552, D-64220 Darmstadt, Germany}
\maketitle
\begin{abstract}
We consider the partition functions of QCD and QED at high temperature
assuming small coupling constants, and present arguments in
favor of an improved perturbative expansion in terms of unstable
excitations. Our effective propagators are derived from spectral
functions with a constant width. These spectral functions
describe screening and damping of gluons (photons) as well as
``Brownian'' motion of quarks (electrons). BRST-invariance allows
us to reduce the number of independent width parameters to three.
These are determined in a self-consistent way from the one-loop self
energy and polarization tensor in the infrared limit thus rendering
this limit finite. All spectral width parameters are found to be
proportional to $g T$. We reproduce the well known expression for the
electric ``Debye''-screening mass.  The transverse (magnetic) gluons
(photons) are found to interact only at nonzero momentum or energy, at
least to leading order. As a consequence their spectral function
acquires a width only away from the infrared limit. Finally, plasmon
modes are determined and found to be strongly damped.
\end{abstract}
\clearpage
\section{Introduction}
It is well known that naive perturbation theory for gauge theories at
finite temperature does not work. This is due to severe infrared
divergences related to the approximation of massless gauge bosons
having an infinite correlation length. The breakdown
of perturbation theory can be understood
qualitatively if one recalls that the average distance between
``particles'' in a heat bath is of order $1/T$, the inverse
temperature. On the other hand, due to the uncertainty relation, each
``particle'' has a quantum mechanical correlation length $1/M$, where
$M$ is its mass. This implies that the individual wave functions
substantially overlap as soon as $T>M$. For massless fields this
condition is met already at infinitesimally small temperature.

As an example providing a more concrete condition consider the gluon self
energy at $\omega=0, \vec{p}\to 0$. The $l$-loop contribution can be
estimated as  \cite[p.145]{Kap89} (Lorentz indices are suppressed)
\begin{equation}
\Pi (\omega=0, \vec{p}\to 0) \approx g^4 T^2\,
  \left(\frac{g^2 T}{M}\right)^{l-2}
\label{est1}
\end{equation}
where $l>2$, $g$ is the coupling constant and $M$ is a fictitious mass
of the gluons serving as an infrared cutoff. Here, one can see the
infrared divergences to become more and more severe the higher the
loop order is.

There exists at least a partial solution to the infrared problem. This
is the method of ``hard thermal loops'' \cite{BP90}. The essence of
this method is the introduction of thermal screening masses. For
electric gluons one finds a mass of the order of $g T$. This suggests
that the loop terms in \fmref{est1} are suppressed by factors of
$g^{l-2}$ as compared to the leading terms. Unfortunately, one finds
no mass of this order for the magnetic or transverse gluons. This
seems to imply that QCD at high temperatures finally is highly
nonperturbative \cite{L80,Br94}.

With the present work we are presenting a slightly modified approach
which may overcome these problems and possibly will allow for an
expansion in the coupling constant. The main difference to
conventional approaches is that we do not rely on quasi-particles,
i.e., particles that have a definite mass-shell only slightly perturbed
by the interactions. There are several arguments against the quasi-particle
concept at finite temperature:
\begin{enumerate}
\item
A system at finite temperature does not exhibit the
Poincar\'e symmetry of the vacuum state. What remains
instead is a semi-direct product of the four-dimensional
translation group T(4) and  the three-dimensional rotation group SO(3).
The irreducible representations of this group have a
continuous mass spectrum \cite{BS75}, and only these
representations may be used in a perturbative expansion \cite{L88}.
\item
A self-consistent calculation of the spectral width
of {\em massive\/} fermions at nonzero temperature $T$ exhibits
a nonanalytical behavior in $T$ as $T \to 0$ \cite{hsw94}.
This is consistent
with the picture of Poincar\'e symmetry restoration at $T=0$ as a
``thermal phase transition''. It also tells us, that the basis
for a perturbative expansion at $T=0$ is different from the basis
at $T>0$ where we have to use a continuous mass spectrum
\cite{L88,habil}.
\item
Consequences of the KMS-condition \cite{KMS,LW87} on quasi-particles have
been investigated by Narnhofer, Requardt and Thirring
\cite{NRT83}. Their theorem essentially states that quasi-particles
can be used as an approximation only if one considers length and time
scales larger that $1/T$, i.e., the regime where the presence of
thermal excitations is still negligible.
\item
Probably the most severe objection against the use of
quasi-particles arises in the long-time limit, i.e., for times larger
than $\approx 1/T$. For such long times the assumption of a
quasi-particle with an infinite life-time no longer holds because the
presence of thermal excitations necessarily leads to scattering or
absorption. Attempts to calculate the damping rates in a perturbative
manner from the imaginary part of the self energy or polarization {\em
on-shell\/} must fail if the mass of the quasi-particle is smaller
than $T$. Indeed, calculations of damping rates indicate that one has
to determine such rates in a self-consistent manner,
 although one frequently makes the assumption that the damping rate is
much smaller than the mass of the quasi-particle \cite{SL90,BK94,hsw94}
\end{enumerate}
It is this breakdown of the quasi-particle picture for temperatures
higher than the relevant masses on which we base our procedure.
We are taking this feature into account establishing simple ans\"atze
for spectral functions with a finite width. This will be the content
of section 2. In section 3 we calculate the self energies and
polarizations to one-loop order
and determine from them the width parameters in a
self-consistent manner. We show these results as well as those for
plasmon modes and their damping rates for QCD, and in section 4 also
for QED. Section 5 is devoted to a summary and conclusions.
%
\section{Spectral functions and imaginary time Green's functions}
The safest method for the determination of Green's functions uses
dispersion relations. They enforce the proper analyticity in the
complex energy plane by removing unphysical poles from the
physical Riemann sheet \cite{hsw94}.
The central quantity of this procedure is a
spectral function ${\cal A} (E,\vec{p})$ which must be known for real
$E$. For arbitrary complex $E$, the retarded and advanced Green's
functions read
\begin{equation}
S^{R,A} (E,\vec{p}) = \int_{-\infty}^\infty \!dE' \; {\cal A}
(E',\vec{p})\, \frac{1}{E-E'\pm{\mathrm i} \epsilon}~~.
\label{dr1}
\end{equation}
This automatically implies that the spectral function is proportional
to the imaginary part of the retarded respectively the advanced
Green's function:
\begin{equation}
{\cal A} (E,\vec{p}) = \mp \frac{1}{\pi} \mbox{Im} (S^{R,A} (E,\vec{p})) =
\frac{1}{2\pi \i} \left( S^{A} (E,\vec{p}) - S^{R} (E,\vec{p}) \right)
\label{dr2}
\;.\end{equation}
For perturbative calculations of the partition function one usually
needs {\em causal Green's functions in euclidean time} respectively
their Fourier transform $S(p_0, \vec{p})$ with the Matsubara
frequencies \cite{M55}
\begin{equation}
p_0 = \left\{ {\array{lcll}
\nu^+_n &=& 2n\;\pi T  &\hbox{(bosons)}\cr
\nu^-_n &=& (2n+1)\; \pi T &\hbox{(fermions)}\endarray}\right.
\;, n = 0, \pm 1, \dots
\label{nupm}
\;.\end{equation}
These functions can be
obtained from the retarded or advanced Green's functions
by analytic continuation in a unique manner \cite{BM61,FW71}, using
the relations
\begin{eqnarray}
\vphantom{\int}
S^{R} (E,\vec{p}) &=& \pm S(p_0 \to \epsilon - {\mathrm i} E,
\vec{p}),~~\epsilon\to 0^+\cr
\vphantom{\int}
S^{A} (E,\vec{p}) &=& \pm S(p_0 \to -\epsilon - {\mathrm i} E, \vec{p})
\label{ac}
\;.\end{eqnarray}
With these relations we can now define $S(p_0, \vec{p})$ via
dispersion integral:
\begin{equation}
S(p_0, \vec{p}) = \mp \int_{-\infty}^\infty \!dE \; {\cal A}
(E,\vec{p})\, \frac{1}{E-{\mathrm i} p_0}
\label{dr3}
\;.\end{equation}
The upper sign in \fmref{ac} and \fmref{dr3} is valid for fermions,
the lower one for bosons; they account for the signs when
switching to euclidean space-time and euclidean $\diGamma$-matrices.

We are interested in spectral functions with a finite {\em constant\/} width.
For bosons, we abbreviate the spectral width as $\gamma_B$ for the moment,
and use the spectral function
\begin{eqnarray}
{\cal A}_B (E,\vec{p}) &=& \frac{1}{\pi} \frac{2 E \gamma_B}{(E^2 -
\gamma_B^2 - \omega_p^2 )^2 + 4 \gamma_B^2 E^2 }\cr
&=&\frac{1}{4 \pi {\mathrm i} \omega_p} \left( \frac{1}{E-\omega_p - {\mathrm
i} \gamma_B} -
\frac{1}{E-\omega_p +{\mathrm i} \gamma_B}\right.\cr
&&\hphantom{\frac{1}{4 \pi {\mathrm i} \omega_p}}\left. +
 \frac{1}{E+\omega_p +{\mathrm i} \gamma_B} -
\frac{1}{E+\omega_p -{\mathrm i} \gamma_B} \right)
\label{sfb}
\;,\end{eqnarray}
where $\omega_p$ denotes the energy-momentum dispersion relation. We
will neglect masses and simply choose $\omega_p^2 = \vec{p}^2$. The
spectral function \fmref{sfb} is normalized according to the
commutation relations for bosonic fields:
\begin{equation}
\int_{-\infty}^\infty \!dE' \; E'\, {\cal A} (E',\vec{p}) = 1
\;.\end{equation}
The corresponding euclidean Green's function can now be obtained using
eqns. \fmref{dr3} and \fmref{sfb}:
\begin{equation}
S_B(p_0, \vec{p}) = \frac{1}{(|p_0| + \gamma_B)^2 + \vec{p}^2 }
\label{sb}
\;.\end{equation}
In the limit of vanishing width, $\gamma_B \to 0$, one recovers the
usual perturbative Matsubara Green's function
$1/(p_0^2+\vec{p}^2 )$.
A remarkable feature is that the Green's function
\fmref{sb} is {\em real}, even with $\gamma_B > 0$.

Similarly, we proceed for fermions, where we abbreviate
the spectral width as $\gamma_F$ to distinguish it from
the Dirac $\diGamma$-matrices $(\diGamma_0,\vec{\diGamma})$.
As spectral function we take
\begin{eqnarray}
{\cal A}_F (E,\vec{p}) &=& \frac{\gamma_F}{\pi}\,
  \frac{\diGamma_0 (E^2 + \omega_p^2 + \gamma_F^2) + {\mathrm i}
  2E\, \vec{\diGamma}\vec{p} }{
(E^2 - \omega_p^2 - \gamma_F^2)^2 + 4E^2\, \gamma_F^2 }\cr
&=&\frac{1}{4 \pi {\mathrm i} \omega_p} \left(
\frac{\diGamma_0\omega_p + {\mathrm i} \vec{\diGamma}\vec{p}}{E-\omega_p
-{\mathrm i} \gamma_F} -
\frac{\diGamma_0\omega_p +{\mathrm i} \vec{\diGamma}\vec{p}}{E-\omega_p +
{\mathrm i} \gamma_F}
\right.\cr
&&\hphantom{\frac{1}{4 \pi {\mathrm i} \omega_p}} \left.
-\frac{\diGamma_0\omega_p-{\mathrm i} \vec{\diGamma}\vec{p}}{E+\omega_p
+{\mathrm i} \gamma_F}
+\frac{\diGamma_0\omega_p-{\mathrm i} \vec{\diGamma}\vec{p}}{E+\omega_p
-{\mathrm i} \gamma_F}
\right)
\label{sff}
\;,\end{eqnarray}
with euclidean $\diGamma$-matrices, see appendix A.
This spectral function is normalized such that
\begin{equation}
\int_{-\infty}^\infty \!dE' \; \mbox{Tr}
\left[\diGamma_0\, {\cal A} (E',\vec{p})\right] = 4~~.
\end{equation}
The corresponding Matsubara Green's function with euclidean
$\diGamma$-matrices and $\omega_p^2  = \vec{p}^2$ simply reads
\begin{equation}
S_F(p_0, \vec{p}) = -{\mathrm i} \;\frac{\diGamma_0\, \mbox{sign} (p_0)\,
(|p_0| + \gamma_F) + \vec{\diGamma}\vec{p}}{
(|p_0| +\gamma_F)^2 + \vec{p}^2 }
\label{sf}
\;.\end{equation}
Obviously, the only modification as compared to the free case is a
shift of the absolute value of the Matsubara frequencies by the
spectral width parameter $\gamma_F$.

For illustration let us consider the spatial and temporal behavior of
a retarded real-time propagator as defined in \fmref{dr1}. With a
boson spectral function according to \fmref{sfb} we get
\begin{eqnarray}
S_B^R (t,\vec{p})&=&\int_{-\infty}^\infty \!dE \; \exp(-{\mathrm i} E t)\,
 S_B^R(E,\vec{p}) \cr
&\propto &
\Theta (t) \exp(-\gamma_B t)\, \frac{\sin (\omega_p t)}{\omega_p}
\label{realt}
\;,\end{eqnarray}
whereas the $\diGamma_0$ component of the fermion propagator with spectral
function \fmref{sff} becomes
\begin{eqnarray}
\mbox{Tr}\left[\diGamma_0 S_F^R (t,\vec{p})\right]
& =& \int_{-\infty}^\infty \!dE \; \exp(-{\mathrm i} E t)\,
\mbox{Tr}\left[\diGamma_0 S_F (E,\vec{p})\right] \cr
&\propto& \Theta (t) \exp(-\gamma_F t) \, \cos( \omega_p t)
\label{realf}
\;.\end{eqnarray}
This shows that $\gamma$ describes the damping of the corresponding
wave-function. In order to investigate the spatial behavior we
consider the time average of \fmref{realt}:
\begin{equation}
\int_{-\infty}^\infty \!dt \; S_B^R (t, \vec{p}) = -{\mathrm i} S_B (0,\vec{p})
=
\frac{-\i}{\vec{p}^2 +\gamma_B^2 }
\label{sstat}
\;.\end{equation}
Obviously, for these zero frequency bosonic modes one cannot distinguish
between a mass and a spectral width. This implies that a spectral
width not only describes damping but also screening.

Another interpretation of these spectral functions, which is
easier to accept in the fermionic case, is the one
of {\em Brownian motion\/} \cite[eq. (6.33)]{CL83}. In this case,
\fmref{realt} resp. \fmref{realf} represent the decaying probability amplitude
to find a particle with momentum $\vec{p}$ in the same state
after time $t$, when it is subject to thermal fluctuations
of the surrounding medium.
%
\section{The partition function of hot QCD}
We now consider the partition function of QCD represented as a
functional integral,
\begin{equation}
Z = \int\! {\cal D}A_\mu {\cal D}\bar{\psi} {\cal D}\psi {\cal
D}\bar{c} {\cal D}c \; \exp (-{\cal S}_{QCD})~~.
\label{pfQCD}
\end{equation}
For convenience we consider a family of $O(3)$-covariant gauges
\cite{Nad83}. As we are interested in high temperatures we neglect
current quark masses. The action reads
\begin{eqnarray}
{\cal S}_{QCD} = \int\!d^4x &\biggl[ & \bar{\psi}_{l,\alpha} (x)
\diGamma_\mu \left(\partial_\mu \delta_{\alpha \beta} - i g A_\mu^a (x)
T^a_{\alpha \beta} \right)\psi_{l,\beta} (x) + \frac{1}{4} F_{\mu\nu}^a
F_{\mu\nu}^a \cr
&& +\frac{1}{2} (\Lambda_\mu A_\mu^a (x))^2 + \Lambda_\mu \bar{c}^a
(x) \left( \partial_\mu c^a (x) + g f^{abc} A_\mu^b (x) c^c (x)
\right) \biggr] ~~~.
\label{SQCD}
\end{eqnarray}
Here, $l,m$ are flavor indices, $\alpha,\beta$ are color indices and
the field strength tensor is defined as
\begin{equation}
F_{\mu\nu}^a = \partial_\mu A_\nu^a - \partial_\nu A_\mu^a + g f^{abc}
A_\mu^b A_\nu^c~~.
\end{equation}
The gauge fixing involves a ``vector'' $\Lambda_\mu$ defined as
\begin{equation}
\Lambda_\mu = (\lambda \partial_\tau ,\frac{1}{\sqrt{\xi}}
\vec{\partial} )
\end{equation}
with gauge parameters $\xi$ and $\lambda$.
The usual perturbative 2-point Green's functions are defined as follows:
\begin{eqnarray}
\langle \psi_{l,\alpha} (x) \bar{\psi}_{m,\beta} (y) \rangle_0 &=&
\delta_{lm} \delta_{\alpha \beta} \int\!\frac{d^4p^-}{(2\pi)^4} S_{F0}
(p)\, \exp({\mathrm i} p (x-y))\cr
\vphantom{\int}
S_{F0} (p) &=& \left({\mathrm i} \diGamma_0 p_0  + {\mathrm i}
\vec{\diGamma}\vec{p}
   \right)^{-1}
\label{defg2a}
\;\end{eqnarray}
for the quarks,
\begin{eqnarray}
\langle A_\mu^a (x) A_\nu^b (y) \rangle_0&=&
 \delta^{ab} \int\!\frac{d^4p^+}{(2\pi)^4}
   \,\exp({\mathrm i} p (x-y)) \cr
\vphantom{\int}
&&\left( S_{T0} (p) {\cal P}^T_{\mu\nu} +
  S_{E0} (p){\cal P}^E_{\mu\nu} + S_{L10} (p) {\cal P}^{L1}_{\mu\nu} +
  S_{L20} (p) {\cal P}^{L2}_{\mu\nu} \right)\cr
\vphantom{\int\limits_1^1}
  S_{T0} (p) &=& \frac{1}{p_0^2 + \vec{p}^2 }\cr
\vphantom{\int\limits_1^1}
  S_{E0} (p) &=& \frac{p_0^2
  +\frac{1}{\xi}\vec{p}^2}{ [\lambda p_0^2 + \frac{1}{\sqrt{\xi}}
  \vec{p}^2 ]^2}\cr
\vphantom{\int\limits_1^1}
S_{L10} (p) &=& \frac{\lambda^2 p_0^2 + \vec{p}^2}{[\lambda
  p_0^2 + \frac{1}{\sqrt{\xi}} \vec{p}^2 ]^2}\cr
\vphantom{\int\limits_1^1}
S_{L20} (p) &=& \left( 1-\frac{\lambda}{\sqrt{\xi}}\right)
  \frac{p_0 \,|\vec{p}|}{[\lambda p_0^2 + \frac{1}{\sqrt{\xi}}
  \vec{p}^2 ]^2}
\label{defg2b}
\end{eqnarray}
for the gluons and
\begin{eqnarray}
\langle c^a (x) \bar{c}^b (y) \rangle_0 &=&
\delta^{ab} \int\!\frac{d^4p^+}{(2\pi)^4} S_{gh0} (p)
 \exp({\mathrm i} p (x-y))\cr
\vphantom{\int}
S_{gh0} (p) &=& \frac{1}{\lambda p_0^2 + \frac{1}{\sqrt{\xi}}
  \vec{p}^2}
\label{defg2c}
\end{eqnarray}
for the ghost fields.
The symbol $\langle \ldots \rangle_0$ denotes functional integration
with measure $\exp (-{\cal S}_{0})$, taking into account only the
bilinear part of the action. The integration over $d^4p^\pm$ is defined in
appendix A. Recall, that for
the corresponding objects in operator formalism time ordering is
implied.

In \fmref{defg2b} we have introduced four symmetric,
$O(3)$-covariant tensors,
\begin{eqnarray}
\vphantom{\int}
{\cal P}^T_{\mu\nu} &=& \delta_{\mu i}(\delta_{ij} - \frac{p_i
   p_j}{\vec{p}^2}) \delta_{j\nu}\cr
\vphantom{\int}
{\cal P}^E_{\mu\nu} &=& \delta_{\mu 0} \delta_{0\nu}\cr
\vphantom{\int}
{\cal P}^{L1}_{\mu\nu} &=& \delta_{\mu i} \frac{p_i
   p_j}{\vec{p}^2} \delta_{j\nu}\cr
\vphantom{\int}
{\cal P}^{L2}_{\mu\nu} &=& \delta_{\mu 0} \frac{p_i}{|\vec{p}|}
\delta_{i\nu} + \delta_{\mu i} \frac{p_i}{|\vec{p}|}
\delta_{0\nu}
\label{tens}
\;.\end{eqnarray}
Among these, ${\cal P}^T, {\cal P}^E$ and ${\cal P}^{L1}$ are
projectors with
\begin{equation}
{\cal P}^T + {\cal P}^E +{\cal P}^{L1} = 1_4~~.
\end{equation}
The square of the traceless tensor ${\cal P}^{L2}$ is the projector
onto the ``longitudinal'' subspace:
\begin{equation}
({\cal P}^{L2})^2 = {\cal P}^E +{\cal P}^{L1}
\;.\end{equation}
Our aim now is to provide propagators having finite spectral widths.
These spectral widths are assumed to be constant and will be
determined self-consistently from the lowest order self energies and
polarizations at zero momentum and energy. This means that we
separate the unperturbed part of the action from the interaction part
in the following way:
\begin{eqnarray}
\vphantom{\int}
{\cal S}_{QCD} &=& {\cal S}_0 + {\cal S}_{int}\cr
\vphantom{\int}
&=& {\cal S}_0 +{\cal S}_\gamma + {\cal S}_{int} - {\cal S}_{\gamma}\cr
\vphantom{\int}
&=& \tilde{\cal S}_0 + \tilde{\cal S}_{int}\cr
\vphantom{\int}
\tilde{\cal S}_0 &=&  {\cal S}_0 +{\cal S}_\gamma
\label{SQCDtilde}
\;.\end{eqnarray}
Perturbation theory is now done with respect to
$\tilde{\cal S}_{int}$, and $\tilde{\cal S}_0$ is the action
of {\em generalized free fields\/}, which already contain the full
two-point correlations present in the theory. In particular, these
generalized free fields may have a continuous mass spectrum,
and therefore avoid the pitfalls of the quasi-particle picture
we outlined in the introduction. This method
has been introduced by Licht \cite{L65}, its application to
systems at finite temperature is discussed in refs. \cite{L88,habil}.

As this procedure constitutes a certain approximation scheme we have
to worry about gauge dependence.
The appropriate tool in this context are BRST-Ward
identities. After gauge fixing the partition function of QCD has
a global symmetry with Grassmann character, the so-called
BRST-symmetry \cite{BRS}:
\begin{eqnarray}
\vphantom{\int}
  \delta \bar{\psi}_{i,\alpha} (x) &~=~& -{\mathrm i} \bar{\psi}_{i,\beta} (x)
  T^a_{\beta ,\alpha} c^a(x) \cr
\vphantom{\int}
  \delta \psi_{i,\alpha} (x) &~=~& {\mathrm i} c^a (x) T^a_{\alpha ,\beta}
  \psi_{i,\beta} (x) \cr
\vphantom{\int}
  \delta A_\mu^a (x) &~=~& \frac{1}{g} \partial_\mu c^a (x) + f^{abc}
  A_\mu^b (x) c^c (x) \cr
\vphantom{\int}
  \delta c^a (x) &~=~& -\frac{1}{2} f^{abc} c^b (x) c^c (x) \cr
\vphantom{\int}
  \delta \bar{c}^a (x) &~=~& \frac{1}{g} \Lambda_\mu A_\mu^a (x)
\label{BRS}
\;.\end{eqnarray}
This symmetry allows to derive the following Ward identity:
\begin{eqnarray}
\vphantom{\int}
\langle \delta [ g A_\mu^a (x) \bar{c}^b (y) ]\rangle &=& 0\cr
\vphantom{\int}
&=&\langle A_\mu^a (x) \Lambda_\nu A_\nu^b (y) + \partial_\mu c^a
(x)\bar{c}^b (y) + g f^{acd} A_\mu^c (x) c^d (x) \bar{c}^b
(y)\rangle\cr
&&
\label{Wi1}
\;.\end{eqnarray}
At leading order in an expansion in powers of the coupling constant
this identity implies the following relations among the perturbative
propagators, which also hold for the effective propagators
of the generalized free fields:
\begin{eqnarray}
\lambda\, p_0 S_{E} (p) + \frac{1}{\sqrt{\xi}} |\vec{p}| S_{L2} (p)
&=& p_0 S_{gh} (p)\cr
\lambda p_0 S_{L2} (p) + \frac{1}{\sqrt{\xi}} |\vec{p}| S_{L1} (p)
&=& |\vec{p}| S_{gh} (p)
\label{Wi2}
\end{eqnarray}
These identities are fulfilled by the expressions in
\fmref{defg2a} -- \fmref{defg2c} as one can see by insertion.

In order to simplify calculations we choose $\lambda = \sqrt{\xi}$
in the following. In use of the simple expressions we derived
in section 2 for propagators with a constant spectral width,
we now provide the following improved propagators for QCD:
\begin{eqnarray}
\vphantom{\int\limits_1^1}
S_{F} (p) &=& \left(
  \vphantom{\int}
  {\mathrm i} \diGamma_0\; \mbox{sign} (p_0)(|p_0| +\gamma_F)  +
  {\mathrm i} \vec{\diGamma}\vec{p} \right)^{-1} =
  \diGamma_\mu S_{F,\mu} (p)\cr
\vphantom{\int\limits_1^1}
S_{T} (p) &=& \frac{1}{(|p_0| +\gamma_T)^2 + \vec{p}^2 }\cr
\vphantom{\int\limits_1^1}
S_{E} (p) &=& \frac{1}{\xi}\frac{1}{(|p_0| +\gamma_E)^2 +
\frac{1}{\xi} \vec{p}^2}\cr
\vphantom{\int\limits_1^1}
S_{L1} (p) &=& \frac{1}{(|p_0| +\gamma_E)^2 + \frac{1}{\xi} \vec{p}^2}\cr
\vphantom{\int\limits_1^1}
S_{L2} (p) &=& 0\cr
\vphantom{\int\limits_1^1}
S_{gh} (p) &=& \frac{1}{\sqrt{\xi}} \frac{1}{(|p_0| +\gamma_E)^2 +
\frac{1}{\xi} \vec{p}^2}
\label{prop}
\;.\end{eqnarray}
As these propagators describe generalized free fields
\cite{L65}, they fulfill the tree-level BRST identities \fmref{Wi2}
{\em by construction\/}. We consider this a crucial point of the present
paper, and emphasize the importance of the BRST symmetry over
particularities of the approximation scheme: From eqn.
\fmref{SQCDtilde} follows, that the identities \fmref{Wi2}
should hold in every order of the {\em resummed\/} perturbation
expansion in powers of $\tilde{\cal S}_\gamma$.

We now have to determine the three unknown parameters $\gamma_F,
\gamma_T$ and $\gamma_E$. For that purpose we calculate the self
energy of the quarks and the polarization tensor for the gluons to
one-loop order {\em with the effective propagators\/}. As is
known from the exact integral equations for Green's functions,
we therefore neglect only three-point and higher correlations.
Up to this level however, our approximation scheme is
completely consistent.

As the parameters $\gamma_F, \gamma_T$ and $\gamma_E$
are due to interactions they are necessarily proportional to some
positive power of the coupling constant $g$ times the temperature $T$.
At temperatures high enough in order to ensure a small coupling
constant the width parameters are necessarily small compared to the
temperature. Therefore, we consider the limit of high temperature (for
details see appendix B). As
we want to provide a well-defined infrared limit with our
parameterization of the spectral functions we determine $\gamma_F,
\gamma_T$ and $\gamma_E$ precisely in this limit.
\begin{figure}[t]
\setlength{\unitlength}{1mm}
\begin{picture}(150,110)
\put(5,95){\large$\Pi(p_0,{\boldsymbol p}) =$}
\put(75,95){\large +}
\put(15,55){\large +}
\put(82,55){\large +}
\put(5,20){\large$\Sigma^F(p_0,{\boldsymbol p}) =$}
\includegraphics{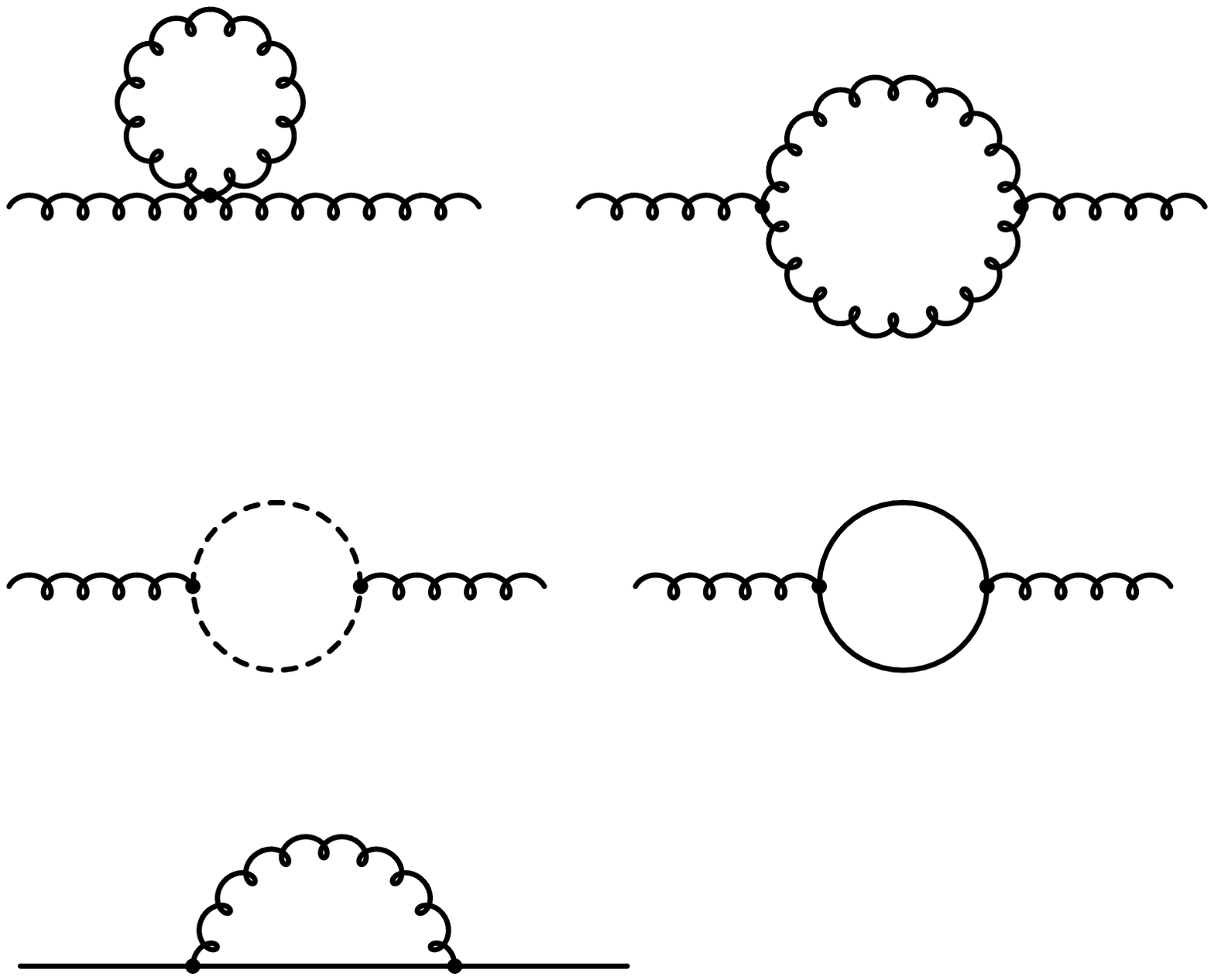}
\end{picture}\\
\caption{Feynman diagrams calculated in this work.}

\hrule
\label{fig1}
\end{figure}

Let us start with the fermions. Our ansatz \fmref{prop} for the
propagators yields the equation
\begin{equation}
\gamma_F = \frac{1}{4} \mbox{Tr} \left[ \diGamma_0
  \Sigma^F(p_0,0) \right]_{p_0\rightarrow 0}~~.
\label{eqF}
\end{equation}
The expression for the one-loop self energy $\Sigma^F (p_0,\vec{p})$ is,
according to diagrammatic rules for the bottom diagram in figure
\ref{fig1},
\begin{equation}
\Sigma^F (p) = {\mathrm i} g^2 \frac{N_c^2 -1}{2 N_c} \int\!\frac{d^4 k}{(2
\pi)^4}\;
S_F (p+k) \left(2 S_T(k) + S_L (k) - S_E (k)\right)~~.
\end{equation}
For high temperatures and for positive $p_0$ we find,
using the procedure outlined in appendix B:
\begin{equation}
\frac{1}{4} \mbox{Tr} \left[ \diGamma_0 \Sigma^F(p_0,0) \right] =
\frac{g^2 T^2}{16} \frac{N_c^2 -1}{N_c} \frac{1}{(p_0 + \gamma_T +
\gamma_F)} + {\cal O}(T)
\label{selfF}
\;.\end{equation}
For the electric and magnetic gluons the equations determining
$\gamma_E$ and $\gamma_T$ are
\begin{equation}
\xi\,\gamma^2_E = \left. \Pi_{00}(p_0,0)\right|_{p_0\rightarrow 0}
\label{eqE}
\end{equation}
and
\begin{equation}
\gamma_T^2 = \left. \frac{1}{2} P^T_{\mu\nu} \Pi_{\mu\nu}(p_0,0)
    \right|_{p_0\rightarrow 0}
\label{eqT1}
\;.\end{equation}
The gluon polarization tensor has a rather complicated structure. We
therefore show only the electric and transverse part which we need
for \fmref{eqE} and \fmref{eqT1}; the following
integrals correspond to the gluon tadpole, gluon/ghost loop and
fermion loop diagrams depicted in figure \ref{fig1}:
\begin{eqnarray}
\vphantom{\int}
\vphantom{\int\limits_0^0}
\Pi_E (p_0, 0) &=\; \Pi_{00} (p_0,0) &\cr
&= \hfill - 2 g^2 N_c \int\!\frac{d^4 k}{(2 \pi)^4} &
  S_T(k_0,\vec{k}) \cr
&\hfill + g^2 N_c \int\!\frac{d^4 k}{(2 \pi)^4} &
  \left[\vphantom{\int}\xi^2 (2 k_0
  \gamma_E + \gamma_E^2 ) S_E (k_0,\vec{k}) S_E (p_0+k_0,\vec{k})\right.\cr
&&\left.\vphantom{\int}
  +  4 (2 k_0 \gamma_T + \gamma_T^2 + \vec{k}^2) S_T (k_0,\vec{k}) S_T
  (p_0+k_0,\vec{k})\right] \cr
&\hfill -2 g^2 N_f  \int\!\frac{d^4 k}{(2 \pi)^4}&
  \left[ S_{F,0} (k_0,\vec{k})\vphantom{\int}
  S_{F,0} (p_0+k_0,\vec{k})\right. \cr
&&\left.\vphantom{\int}
 - S_{F,i} (k_0,\vec{k}) S_{F,i}
(p_0+k_0,\vec{k}) \right]
\;,\end{eqnarray}
\begin{eqnarray}
\Pi_T (p_0, 0) &=\; \frac{1}{2} P^T_{\mu\nu} \Pi_{\mu\nu}(p_0,0) \cr
&= \hfill g^2 N_c \int\!\frac{d^4 k}{(2 \pi)^4}&
  \left[ \vphantom{\int}
  \xi\frac{1+ \cos^2 (\theta )}{2} S_E (k_0,\vec{k}) +
\frac{3-\cos^2 (\theta )}{2} S_T (k_0,\vec{k})\right] \cr
&\hfill + g^2 N_c \int\!\frac{d^4 k}{(2 \pi)^4} &
  \left[\vphantom{\int} \xi (2 \gamma_E k_0 +
  \gamma_E^2 - 4 p_0 k_0 - 4 p_0^2 ) \frac{1- \cos^2 (\theta )}{2}
  \right.  \cr
&&\;\;\;\;\;\;\; \cdot S_E (k_0,\vec{k}) S_E (p_0+k_0,\vec{k})
  \vphantom{\lim_{LLL}} \cr
&& + (2 \gamma_T (k_0 + p_0) +
  \gamma_T^2 - 2 k_0 p_0 - 3 p_0^2) \frac{1+ \cos^2 (\theta )}{2}
  \cr
&&\;\;\;\;\;\;\; \cdot  S_E(k_0,\vec{k}) S_T (p_0+k_0,\vec{k})
  \vphantom{\lim_{LLL}} \cr
&&  \left.\vphantom{\int} + 4 \vec{k}^2 \frac{1- \cos^2
  (\theta )}{2}  S_T (k_0,\vec{k}) S_T (p_0+k_0,\vec{k}) \right]\cr
&\hfill +2 g^2 N_f  \int\!\frac{d^4 k}{(2 \pi)^4}&
  \left[ S_{F,0} (k_0,\vec{k}) S_{F,0} (p_0+k_0,\vec{k})
  \vphantom{\int}\right.\cr
&&\left.\vphantom{\int}
  + S_{F,i} (k_0,\vec{k})\, p_i\, S_{F,j}
  (p_0+k_0,\vec{k})\, p_j/\vec{p}^2\, \right]
\;.\end{eqnarray}
Here we have used the simple relations between $S_E(p), S_L(p)$ and
$S_{gh}(p)$ due to \fmref{Wi2}.
The resulting expressions for high temperature are:
\begin{equation}
\Pi_E (p_0, 0) = -\frac{g^2 T^2 N_c}{6} + \frac{g^2 T^2 N_f
  \gamma_F}{3 (2 \gamma_F + p_0)} + \frac{g^2 T^2 N_c (6 \gamma_T +
  p_0)}{6 (2 \gamma_T + p_0)} + {\cal O}(T)
\label{polE}
\;,\end{equation}
\begin{equation}
\Pi_T (p_0, 0) = p_0 \frac{g^2 T^2}{18} \left(
  \frac{N_f}{2 \gamma_F + p_0} +
  \frac{2 N_c}{2 \gamma_T + p_0} \right)
+ {\cal O}(T)
\label{polT}
\;.\end{equation}
Note, that quite similar results are obtained in a semiclassical
approximation to the bremsstrahlung problem in hot and dense matter
\cite{KV95}. This indicates, that the high temperature limit we consider
is closely related to a classical self-consistent calculation.

Obviously, $\Pi_T (p_0, 0) $ vanishes for $p_0 \to 0$ suggesting
$\gamma_T=0$! On the other hand, it starts linear with $p_0$
supporting a term $2 p_0 \gamma_T$ as in our ansatz for the transverse
gluon propagator. We interpret this behavior as an inconsistency of
our ansatz for the transverse propagator requiring a modification.

The simplest modification of our ansatz
such that $S_T (p_0,0)^{-1}$ vanishes for $p_0 \to 0$ can be
achieved by a spectral function like \fmref{sfb} but now with $\omega_p^2 =
\vec{p}^2 - \gamma_T^2$. The modified transverse propagator then reads:
\begin{equation}\label{propmod}
S_{T} (p) = \frac{1}{|p_0|^2 + 2 |p_0| \gamma_T + \vec{p}^2 }
\;.\end{equation}
It seems that this propagator still has an infrared problem. However,
recall that the proper quantity to consider are correlations of the
field strength tensor rather than the vector potentials. In
particular, the time averaged screening of an external magnetic field
is given by the momentum space correlation function
\begin{equation}
\int\! dt \langle B_i^a (t,\vec{x}) B_j^a (0,\vec{y}) \rangle\;  \approx\;
\int\! d^3 \vec{p}\; \exp (i\vec{p}\vec{x}) {\cal P}^T_{ij} \vec{p}^2 S_T
(0,\vec{p})
\;\propto\; \delta(\vec{x}) ~~~.
\end{equation}
Here, we have not displayed any factors which are irrelevant for the
spatial behavior. Obviously, static magnetic fields are screened and,
in particular, the screening length is even zero!

Doing the previous calculations with the new
spectral function for the transverse gluons yields the same results
for the leading terms. A possible explanation for
this feature is that at high temperature the kinetic terms are
the dominant ones; these are unchanged by our modification.
As the results are the same, our ansatz for the transverse gluons now
is compatible with the infrared limit of the corresponding
transverse polarization. The new equation determining $\gamma_T$
reads:
\begin{equation}
\gamma_T = \left(\frac{1}{4 p_0} P^T_{\mu\nu} \Pi_{\mu\nu}(p_0,0)
     \right)_{p_0\rightarrow 0}
\;.\label{eqT2}
\end{equation}
Now we can solve the set of equations \fmref{eqF},\fmref{eqE} and
\fmref{eqT2} using the expressions \fmref{selfF},\fmref{polE} and
\fmref{polT}. The solutions with positive values for the spectral
width parameters are
\begin{eqnarray}
\vphantom{\int\limits_0^0}
\gamma_E & = & g T \sqrt{ \frac{N_f+ 2 N_c}{6 \xi}}\cr
\vphantom{\int\limits_0^0}
\gamma_T & = & \frac{1}{16}\, \frac{N^2_c - 1}{N_c} \,
               \frac{(g T)^2}{\gamma_F} - \gamma_F\cr
\vphantom{\int\limits_0^0}
\gamma_F & = & g T \frac{1}{12}
               \sqrt{ 11 N_c -\frac{9}{N_c}- N_f -
               \sqrt{(N_f-2 N_c)^2 + 36 (N_c^2-1)}}
\label{gEFT}
\;.\end{eqnarray}
The solution for $\gamma_E$ seems to indicate a gauge dependence.
However, only for $p_0 = 0$ the electric propagator $S_E (p)$ is gauge
invariant. There it reads
\begin{equation}
S_E (0,\vec{p}) = \frac{1}{\vec{p}^2 + \xi \gamma_E^2}~~.
\end{equation}
Obviously, $\xi \gamma_E^2$ is independent of $\xi$ and is just the
famous ``Debye''-screening mass for electric gluons \cite{BP90,Pi64}.
\begin{table}[t]
\begin{center}
\begin{tabular}{|r||c|c|c||c|c|c|}
\hline
 & \multicolumn{3}{|c||}{$N_c = 2$} &\multicolumn{3}{|c|}{$N_c = 3$}
\\ \cline{2-7}
 &$N_f = 1$ &$N_f = 2$ &$N_f = 3$ &$N_f = 1$ &$N_f = 2$& $N_f = 3$
\\ \hline\hline
$\sqrt{\xi}\,\gamma_E/(g T)$ & 0.913 & 1 & 1.080 & 1.080 & 1.155 & 1.225
\\ \hline
$\gamma_T/(g T)$ & 0.273 & 0.323 & 0.390 & 0.315 & 0.344 & 0.380
\\ \hline
$\gamma_F/(g T)$ & 0.199 & 0.185 & 0.168 & 0.28 & 0.271 & 0.260
\\ \hline
\end{tabular}
\end{center}
\caption{Self-consistent spectral width parameters for high temperature QCD}

\hrule
\label{tab1}
\end{table}

Another remarkable feature of these results is the decrease of
$\gamma_F$ with $N_f$. Indeed, for
\begin{equation}
N_f > 9 \frac{N_c^2-1}{2 N_c}
\end{equation}
we get a complex $\gamma_F$. This is not a very serious problem for
the values of $N_f$ and $N_c$ that are realized in nature;
for $N_c=3$ the limit is $N_f<12$. However, this behavior has to be
understood from a purely theoretical point of view as it signals some
strange and unexpected behavior. A possible explanation is that
for large $N_f$ the one-loop diagrams are not the appropriate ones to
describe the leading behavior. In this case, one would have
to consider a large-$N_f$ expansion instead of a loop expansion.

In table \ref{tab1}, we provide a
listing of $\gamma_E,\gamma_T$ and $\gamma_F$ from \fmref{gEFT}
in units of $gT$ for several values of $N_c$ and $N_f$.

For completeness we also mention the results for pure Yang-Mills
theory:
\begin{eqnarray}
\vphantom{\int\limits_0^0}
\gamma_E & = & g T \sqrt{ \frac{N_c}{3 \xi}}\cr
\vphantom{\int\limits_0^0}
\gamma_T & = & g T \frac{\sqrt{N_c}}{6}
\label{gamYM}
\;.\end{eqnarray}
One may also determine the plasma frequency of the system.
We concentrate on the
transverse oscillations,  because only these are gauge invariant.
They are solutions of the equation
\begin{equation}
({\mathrm i} \omega_T)^2 = \Pi_T ({\mathrm i} \omega_T,0)
\;.\end{equation}
There is always one solution with $\omega_T = 0$ because $\Pi_T
(p_0,0)$ is proportional to $p_0$. Numerical values for the other
solutions are listed in table \ref{tab2}.
\begin{table}[t]
\begin{center}
\begin{tabular}{|r||c|c|c|}
\hline
 & \multicolumn{3}{|c|}{$N_c = 2$} \\ \cline{2-4}
 &$N_f = 1$ &$N_f = 2$ &$N_f = 3$ \\
\hline \hline
$\omega_T/(g T)$ & 0.457 - 0.255 ${\mathrm i}$
& 0.496 - 0.263 ${\mathrm i}$ & 0.524 - 0.258 ${\mathrm i}$
\\ \hline
$\omega_T/(g T)$ & -0.434 ${\mathrm i}$ & -0.489 ${\mathrm i}$
& -0.6 ${\mathrm i}$
\\ \hline \hline
&\multicolumn{3}{|c|}{$N_c = 3$}
\\ \cline{2-4}
&$N_f = 1$ &$N_f = 2$& $N_f = 3$ \\
\hline\hline
$\omega_T/(g T)$ &     0.541 - 0.309 ${\mathrm i}$
& 0.579 - 0.323 ${\mathrm i}$ & 0.614 - 0.331 ${\mathrm i}$
\\ \hline
$\omega_T/(g T)$ &  -0.571 ${\mathrm i}$ & -0.584 ${\mathrm i}$
&   -0.617 ${\mathrm i}$
\\ \hline
\end{tabular}
\end{center}
\caption{Transverse plasma frequencies of high temperature QCD}

\label{tab2}
\hrule
\end{table}
A remarkable feature is the strong damping of these modes. Their
imaginary part is about half of the real part. Moreover, there is also
an overdamped mode. The ratio between the electric screening mass
$\sqrt{\xi} \gamma_E$ and the plasmon frequency $\mbox{Re} (\omega_T)$ is
almost exactly $2$ in contrast to the standard value $\sqrt{3}$
\cite{L80,Pi64}. The explanation for this discrepancy may be the
inclusion of an imaginary part of $\omega_T$ which is not negligible.

\section{The partition function of hot QED}
Essentially the same procedure can be applied to QED. Let us briefly
recapitulate the most important steps. The partition function is now
\begin{equation}
Z_{QED} = \int\! {\cal D}A_\mu {\cal D}\bar{\psi} {\cal D}\psi {\cal
D}\bar{c} {\cal D}c \; \exp (-{\cal S}_{QED})
\label{pfQED}
\end{equation}
with the action
\begin{eqnarray}
{\cal S}_{QED} = \int\!d^4x &\biggl[ & \bar{\psi}_l (x)
\diGamma_\mu \left(\partial_\mu - i e A_\mu (x) \right)\psi_l (x) +
\frac{1}{4} F_{\mu\nu} F_{\mu\nu} \cr
&& +\frac{1}{2} (\Lambda_\mu A_\mu (x))^2 + \Lambda_\mu \bar{c}
(x) \partial_\mu c (x) \biggr]
\label{SQED}
\;,\end{eqnarray}
where $l$ is again a ``flavor'' index. The electromagnetic field
strength tensor is
\begin{equation}
F_{\mu\nu}^a = \partial_\mu A_\nu^a - \partial_\nu A_\mu^a
\;,\end{equation}
and the gauge is the same as for QCD.

We now define a modified unperturbed action $\tilde{\cal S}_0$:
\begin{eqnarray}
\vphantom{\int}
{\cal S}_{QED} &=& {\cal S}_0 + {\cal S}_{int}\cr
\vphantom{\int}
&=& {\cal S}_0 +{\cal S}_\gamma + {\cal S}_{int} - {\cal S}_{\gamma}\cr
\vphantom{\int}
&=& \tilde{\cal S}_0 + \tilde{\cal S}_{int}\cr
\vphantom{\int}
\tilde{\cal S}_0 &=&  {\cal S}_0 +{\cal S}_\gamma
\label{SQEDtilde}
\;.\end{eqnarray}
Repeating the same procedure as for QCD means
that we also introduce a spectral width for the ghosts.
One might ask, whether this is not a contradiction to the fact
that the ghosts are free fields that decouple from the
physical fields. However, this question is irrelevant in our
method, since we are dealing with {\em generalized free fields}
having a continuous mass spectrum. Hence, we may choose {\em
any\/} spectral width parameter for unobservable fields \cite{L88}.

Our special choice for the ghost propagator is obtained
by enforcing the BRST identities \fmref{Wi2}. This ensures
the (gauge dependent) choice $S_{L2} (p) =0$, and therefore
is motivated by simplicity of computation.

The equations for $\gamma_F, \gamma_E$ and $\gamma_T$ are the same as
in the QCD-case, namely \fmref{eqF}, \fmref{eqE} and \fmref{eqT2}. The
fermion self energy now is:
\begin{equation}
\Sigma^F (p) = {\mathrm i} e^2 \int\!\frac{d^4 k}{(2 \pi)^4}\;
S_F (p+k) (2 S_T(k) + S_L (k) - S_E (k))~~.
\end{equation}
For high temperature and for positive $p_0$ we find
\begin{equation}
\frac{1}{4} \mbox{Tr} \left[ \Gamma_0 \Sigma^F(p_0,0) \right] =
\frac{e^2 T^2}{8} \frac{1}{(p_0 + \gamma_T +
\gamma_F)} + {\cal O}(T)
\label{SigF}
\;.\end{equation}
The electric and transverse component of the photon polarization
tensor read:
\begin{eqnarray}
\Pi_E (p_0, 0) &=& \Pi_{00} (p_0,0) \cr
&=& -4 e^2 N_f  \int\!\frac{d^4 k}{(2 \pi)^4}\; \left[ S_{F,0} (k_0,\vec{k})
S_{F,0} (p_0+k_0,\vec{k}) \vphantom{\int}\right.\cr
&&\;\;\;\;\;\;\;\;\;\;\;\;\;
\left.\vphantom{\int} - S_{F,i} (k_0,\vec{k}) S_{F,i}
(p_0+k_0,\vec{k}) \right]
\;,\end{eqnarray}
\begin{eqnarray}
\Pi_T (p_0, 0) &=& \frac{1}{2} P^T_{\mu\nu} \Pi_{\mu\nu}(p_0,0) \cr
&=& 4 e^2 N_f  \int\!\frac{d^4 k}{(2 \pi)^4}\; \left[ S_{F,0} (k_0,\vec{k})
S_{F,0} (p_0+k_0,\vec{k})\vphantom{\int}\right.\cr
&&\;\;\;\;\;\;\;\;\;\;\;\;\;
\left.\vphantom{\int} + S_{F,i} (k_0,\vec{k}) p_i S_{F,j}
(p_0+k_0,\vec{k}) p_j/\vec{p}^2 \right]
\;.\end{eqnarray}
The resulting expressions for high temperature are:
\begin{equation}
\Pi_E (p_0, 0) = \frac{2 e^2 T^2 \gamma_F N_f}{3 (2 \gamma_F + p_0)} +
{\cal O}(T)
\label{plE}
\;,\end{equation}
\begin{equation}
\Pi_T (p_0, 0) = \frac{e^2 T^2 N_f p_0}{9 (2 \gamma_F + p_0)}
+ {\cal O}(T)
\label{plT}
\;.\end{equation}
Solving the self-consistency equations for the
spectral width parameters in QED then yields
\begin{eqnarray}
\vphantom{\int\limits_0^0}
\gamma_E & = & e T \sqrt{ \frac{N_f}{3 \xi}}\cr
\vphantom{\int\limits_0^0}
\gamma_T & = & \frac{(e T)^2 N_f}{36 \gamma_F} \cr
\vphantom{\int\limits_0^0}
\gamma_F & = & \frac{e T}{2} \sqrt{\frac{1}{2} - \frac{N_f}{9}}
\label{gamQED}
\;.\end{eqnarray}
Again we find that for $N_f>4$ the fermionic spectral width $\gamma_F$
becomes imaginary (for a discussion see section 3).

Finally, we also give the result for the transverse plasmon mode:
\begin{equation}
\omega_T = e T \frac{1}{2} \sqrt{\frac{5}{9} N_f - \frac{1}{2}} -\i
\gamma_F
\end{equation}
For standard electrodynamics with $N_f=1$ this yields
\begin{equation}
\omega_T = e T \frac{1}{2 \sqrt{18}} (1 -{\mathrm i} \sqrt{7})
  \approx e T \, (0.118 - 0.312 \i)
\end{equation}
This is a strongly damped oscillation. Here, the
ratio between electric screening mass $\sqrt{\xi} \gamma_E$ and
the real part of $\omega_T$ deviates even more strongly from
$\sqrt{3}$ than for QCD. However, since the plasma oscillation is also
stronger dameped than in the QCD case, this only supports our previously
stated conclusion that one may not negelect the imaginary part
for the determination of the plasmon mode.
%
\section{Summary and conclusions}
In the present paper, we have investigated the self-consistent
determination of finite temperature spectral broadening for
QCD and QED in the infrared limit, i.e., for zero ``particle''
momenta. For the ``electric'' part of the gauge boson propagator,
we reproduce the well-known Debye screening mass, whereas for
the fermions as well as for the transverse gauge bosons we
acquire new results: In contrast to the method of hard thermal loops
\cite{BP90} we obtain, that all spectral width
parameters $\gamma_F$, $\gamma_E$, $\gamma_T$
are of the order $gT$, respective $eT$, in the high temperature
limit.

Based on results for {\em massive fermions\/} \cite{hsw94} we believe
that in some regimes the constant spectral width obtained here might
be a good approximation to the full solution of the problem.
However, we have formulated a consistent recipe for the
perturbative expansion of QCD and QED in terms of
{\em generalized free fields\/} \cite{L65},
whose continuous mass spectrum is described by our effective
propagators (eqn. \fmref{prop}, transverse boson propagator
replaced by \fmref{propmod}).

The skeleton expansion using these propagators
may serve to calculate the full QCD and
QED Green's functions \cite{L88,habil}.
Indeed, if the estimate \fmref{est1} holds for
any infrared regulator, $i.e.$ if $M$ may be replaced by a typical
$\gamma$, our results suggest that now a perturbative calculation of
high temperature QCD or QED is possible; the expansion parameter
in this cased would be $g$ instead of $g^2$.

The virtues of a finite temperature perturbation theory
using such propagators with a nontrivial spectral
function lie at hand:
\begin{enumerate}
\item
It takes into account the Narnhofer-Thirring theorem described
in the introduction of our paper \cite{NRT83}.
\item
It takes into account that
the {\em basis\/} of a perturbative expansion
has to be chosen according to the symmetry of the problem:
The symmetry group of a temperature state is {\em not\/}
the Poincar\'e group with its stable particle representations,
but rather possesses only representations with continuous
mass spectrum \cite{BS75,L88}.
\item
It is {\em free of unphysical infrared divergences\/} in every order,
because products of mass-shell $\delta$-functions cannot appear.
\end{enumerate}

Finally, our results for the transverse plasmon modes indicate a
rather strong damping in contrast to the frequently used assumption
that damping of these collective modes is suppressed by a factor of
$g$ or $e$. Due to the large imaginary part of these plasmon
frequencies the relation $m_{el} = \sqrt{3}\mbox{Re} (\omega_T)$ no longer
holds.

\subsection*{acknowledgement}
The authors wish to thank B.Friman and J.Knoll for fruitful discussions.
\appendix
\section{Notation}
We use a euclidean metric $g_{\mu\nu} = \delta_{\mu\nu}$. Space-time
integrals are denoted by
\begin{equation}
\int\!d^4x = \int\!d^3x \int_0^\beta\!d\tau~~,~~\beta = 1/T~~.
\end{equation}
The euclidean $\diGamma$-matrices are Hermitean and obey the anticommutation
relations
\begin{equation}
\left\{ \diGamma_\mu , \diGamma_\nu \right\} = 2 \delta_{\mu\nu}~~.
\end{equation}
The gluon fields $A_\mu = A_\mu^a T^a$ are local elements of the Lie
algebra SU(N). As such $A_\mu$ is traceless and Hermitean. The
commutation relations of the generators $T^a$ are
\begin{equation}
[T^a,T^b] = {\mathrm i} f^{abc} T^c~~.
\end{equation}
They are normalized such that
\begin{equation}
\mbox{Tr}\left[ T^a T^b \right] = \frac{1}{2} \delta^{ab}
\end{equation}
and their completeness relation reads
\begin{equation}
T^a_{\alpha\beta} T^a_{\rho\sigma} = \frac{1}{2} (\delta_{\alpha\sigma}
\delta_{\rho\beta} - \frac{1}{N} \delta_{\alpha\beta}
\delta_{\rho\sigma}) ~~.
\end{equation}
This implies
\begin{equation}
T^a_{\alpha\beta} T^a_{\beta\sigma} = \frac{N^2 -1}{2 N}
\delta_{\alpha\sigma} ~~.
\end{equation}
A useful relation for the structure constants reads
\begin{equation}
f^{abc} f^{dbc} = N \delta^{ad}~~.
\end{equation}

Fourier transforms of a field $\phi(x)$ are defined as follows:
\begin{eqnarray}
\phi(x) &=& \phi(\tau,\vec{x}) = \int\!\frac{d^4p^\pm}{(2\pi)^4}
\phi(p) e^{{\mathrm i} p x}\cr
&=& \int\!\frac{d^3p}{(2\pi)^3} \frac{1}{\beta} \sum_n \phi(\nu_n^\pm
, \vec{p}) e^{{\mathrm i} \nu_n^\pm \tau + i\vec{p}\vec{x}}~~.
\end{eqnarray}
Here, $\nu_n^\pm$ are the Matsubara frequencies for bosons (fermions)
depending on the character of the field $\phi$, and defined in \fmref{ac}.

\section{High temperature limit}
Sums over Matsubara frequencies are transformed into
contour integrals using the well-known identity:
\begin{equation}
\sum_n f(\nu_n) = \int_{-\infty}^\infty \! \frac{dz}{2 \pi} f(z) \pm
\int_{-\infty+{\mathrm i} \epsilon}^{+\infty+{\mathrm i} \epsilon} \!
\frac{dz}{4 \pi}
\frac{f(z) + f(-z)}{\exp (-{\mathrm i} \beta z) \mp 1}
\end{equation}
The upper sign is for bosonic $\nu_n$, the lower sign for fermionic
$\nu_n$.

The remaining one-loop integrals have the following typical form
(see ref. \cite{habil} for details):
\begin{equation}
\int\limits_{-\infty}^\infty
   \!\!dE\,\frac{1}{\exp\left(\displaystyle \frac{E}{T}\right)
   \pm 1}\,\int\limits_0^\infty\!\!d k\, k^2\,{\cal A}_1(E,k)\,
   \int\limits_{-\infty}^\infty\!\!dE^\prime\,\frac{
   {\cal A}_2(E^\prime,k)}{E-E^\prime+ p_0}
\end{equation}
The $E^\prime$ integration as well as the $k$-integration are carried
out analytically, as contour integrals in the complex plane of the
corresponding integration variable.

For the remaining $E$-integration, we use the fact that the integrand
is an odd function of $E$ depending only on the scale $\gamma$,
\begin{equation}
f(E) = \gamma \,\tilde{f}(\frac{E}{\gamma})~~~,~~~f(E) = -f(-E)
\;.\end{equation}
multiplied by a temperature distribution function. For simplicity we
discuss the case where $f(E)$ has energy-dimension $1$.

The $E$-integration may be split up into two parts,
\begin{equation}
\int\limits_0^\infty \!\!dE\,\frac{f(E)}{\exp\left(
  \displaystyle\frac{E}{T}\right)\pm 1}
=  \gamma T \int\limits_{\gamma/T}^\infty\!\!
  dx\,\frac{\tilde{f}(x \frac{T}{\gamma})}{\exp(
  \displaystyle x) \pm 1}
+  \gamma^2 \int\limits_0^1\!\!
  dx\,\frac{\tilde{f}(x)}{\exp\left(x\displaystyle
  \frac{\gamma}{T}\right) \pm 1}
\;.\end{equation}
In the first part, we may use an approximation for $\tilde{f}$
at {\em large\/} arguments,
\begin{equation}
\tilde{f}\left(x\frac{\gamma}{T}\right) \approx \frac{x T}{\gamma}
+ {\cal O}(\frac{\gamma}{x T})\;.
\end{equation}
The contribution of the leading term to the complete expression is of
order $T^2$. The remainder is at most of order $\gamma T$.

In the second piece of the integrals, we expand the distribution function
for small $x$. The resulting integrals are infrared finite because
$\tilde{f} (x)$ is finite at $x\to 0$. As it is an odd function of its
argument it vanishes at least proportional to $x$ for small $x$. The
leading term arises for bosons where the distribution function behaves
as
\begin{equation}
\frac{1}{\exp \left(x\displaystyle\frac{\gamma}{T}\right) -1} \approx
\frac{T}{\gamma x} + {\cal O}(1)~~~.
\end{equation}
This yields a contribution of order $\gamma T$.


\end{document}